\documentstyle[12pt]{ioplppt}

\begin{document}

\newcommand{\lesssim}{~\raisebox{.6ex}{$<$}\hspace{-9pt}\raisebox{-.6ex}
{$\sim $}~}

\sloppy

\jl{2}

\letter{Strong dependence of multiphoton detachment rates on the asymptotic
behaviour of the ground-state wave function}

\author{G F Gribakin\dag , V K Ivanov\ddag ,  A V Korol\S\
and M Yu Kuchiev\dag\ftnote{4}{E-mails: gribakin@newt.phys.unsw.edu.au,
ivanov@tuexph.stu.neva.ru, Korol@rpro.ioffe.rssi.ru,
kuchiev@newt.phys.unsw.edu.au}}

\address{\dag School of Physics, The University of New South Wales,
Sydney 2052, Australia}

\address{\ddag Department of Experimental Physics, St Petersburg State 
Technical University, Polytekhnicheskaya 29, St Petersburg 195251,
Russia}

\address{\S Physics Department, Russian Maritime Technical
University, Leninskii prospect 101, St Petersburg 198262, Russia}


\begin{abstract}
Two-photon detachment from the $F^{-}$ negative ion has been investigated
within the lowest order perturbation theory. We show that in accordance
with the adiabatic theory a proper asymptotic behaviour of the
$2p$ bound state wave function is crucial for obtaining correct absolute
values of the multiphoton detachment cross sections. We find that the
latter are substantially higher than it was previously believed.

\end{abstract}




\maketitle


In the recent papers by Gribakin and Kuchiev (1997a, 1997b) an adiabatic
analytical theory of multiphoton detachment from negative ions has been
developed, based on the Keldysh approach (Keldysh 1964). Simple
analytical expressions obtained there for the differential and total
$n$-photon detachment cross sections allow one to estimate them for any
negative ion. One of the important points of that work is that the electron
escape from the atomic system in a low-frequency laser field takes place at
large distances,
\begin{equation}\label{large}
r\sim 1/\sqrt{\omega }\sim \sqrt{2n}/\kappa \gg 1,
\end{equation}
where $\omega $ is the photon frequency, $\kappa $ is related to the
initial
bound state energy, $E_0=-\kappa ^2/2$, and $n$ is the number of quanta
absorbed (atomic units are used throughout). Accordingly, the multiphoton
detachment rates are determined by the long-range asymptotic behaviour of
the bound-state wave function, namely by the asymptotic parameters $A$ and
$\kappa $ of the bound state radial wave function
$R(r)\simeq Ar^{-1}e^{-\kappa r}$. This result has been obtained using
the length form of the interaction with the laser field, which proves to be
the most convenient for multiphoton processes.

The analytical adiabatic approach is valid for {\em multiphoton} detachment
processes, i.e., strictly speaking, for $n\gg 1$. However, the calculations
for H$^-$ and halogen negative ions indicate (Gribakin and Kuchiev 1997a,
Kuchiev and Ostrovsky 1998) that the analytical formulae should
give reasonable answers even for $n=2$. The aim of the present work is to
verify these conclusions by doing direct numerical calculations of the
2-photon detachment cross
sections. In particular we examine the sensitivity of the photodetachment
cross sections to the asymptotic behaviour of the ground state wavefunction
and show that it is indeed very strong. Thus, a ``small'' 20\% error
in $\kappa $ present in the Hartree-Fock (HF) wave function of the fluorine
negative ion results in a factor of three underestimation of the 2-photon
cross section. This emphasizes the need to use bound-state wavefunctions
with correct asymptotic behaviour in calculations of multiphoton
processes.

In this paper we calculate the 2-photon detachment amplitudes, cross
sections and photoelectron angular distribution using the lowest-order
perturbation theory and compare the results obtained with different ground
state wavefunctions. We present and analyze the results for the
F$^{-}$ negative ion, where the results of a few other theoretical
calculations (Robinson and Geltman 1967, Crance 1987a,b, Pan \etal 1990,
Pan and Starace 1991, van der Hart 1996) as well as experimental data
(Kwon \etal 1989, Blondel \etal 1992, Blondel and Delsart 1993) are known.
Pan \etal and Pan and Starace calculated the two-photon detachment cross
section and photoelectron angular distribution in the HF approximation
(similar to that used by Crance) and with the account of first-order
electron correlation effects. Their results
show that the correlation corrections are about 15\% for the partial and
total cross sections and almost negligible for the angular distribution
parameters, when the dipole length form is used.


The total cross section of the two-photon detachment of an electron from an
atomic system by a linearly polarized light of frequency $\omega $ is
\begin{equation}\label{cr}
\sigma =\sum _{l_fL} \sigma _{l_fL}=\frac{16\pi ^3}{c^2}\omega ^2
\sum _{l_fL}\left| A_{l_fL}( \omega ) \right| ^2
\end{equation}
where $\sigma _{l_fL}$ is the partial cross section for the detachment into
the final state with the photoelectron orbital momentum $l_f$ and the total
orbital momentum $L$, and the continuous spectrum wave function of
the photoelectron is normalized to the $\delta $-function of energy. For
the $2p$ electron detachment from F$^-~2p^6~^1S$ the final state can be
either $^1S$ ($L=0$, $l_f=1$) or $^1D$ ($L=2$, $l_f=1,~3$). The
two-photon amplitude $A_{l_fL}( \omega )$ is determined by the following
equations
\begin{equation}\label{me1}
A_{l_fL}( \omega ) =\sqrt{2L+1}\left( 
\begin{array}{ccc}
1 & 1 & L \\
0 & 0 & 0
\end{array}
\right)
\sum _l (-1)^l \left\{ 
\begin{array}{ccc}
1 & 1 & L \\ 
l_f & l_0 & l
\end{array}
\right\} M_{l_fl}( \omega )~,
\end{equation}

\begin{equation}\label{me2}
M_{l_fl}( \omega ) =\sum _\nu \frac{\langle \varepsilon _fl_f \|\hat d \|
\nu l \rangle \langle \nu l\| \hat d \| n_0l_0\rangle }
{E_0 +\omega -E_\nu + \i 0 }
\end{equation}
where $\nu l$ is the intermediate electron state with the orbital momentum
$l$ after absorbing the first photon ($l=0,2$ for F$^{-}$), and $n_0l_0$
is the initial bound state. The reduced dipole matrix elements are defined
in the usual way, e.g. in the length form
\begin{equation}\label{dme3}
\langle \nu l\| \hat d \| n_0l_0\rangle =(-1)^{l_>}\sqrt{l_>}
\int P_{\nu l}(r) P_{n_0l_0}(r) r\d r ,
\end{equation}
where $l_>=\mbox{max}\{ l,l_0\}$ and $P$'s are the radial wave
functions.

The photoelectron angular distribution is described by the differential
cross section
\begin{equation}\label{dcr}
\frac{d\sigma }{d\Omega }=\frac \sigma {4\pi }\sum _{j=0}^2
\beta_{2j}( \omega )P_{2j}( \cos \theta ) ~,
\end{equation}
where $\theta $ is measured with respect to the light polarization axis,
and the asymmetry parameters $\beta _{2j}$ are defined in
terms of the two-photon transition amplitudes $A_{l_fL}$ and scattering
phases of the photoelectron $\delta _{l_f}$:
\begin{eqnarray}\label{beta}
\fl \beta _{2j}=\frac{16\pi ^3\omega ^2}{c^2\sigma }(4j+1)\mbox{Re}
\Biggl[ \sum _{l_f^\prime L^\prime l_f^{\prime \prime } L^{\prime \prime }}
(-1)^{l_0+L^\prime +L^{\prime \prime}}
(-\i )^{l_f^\prime +l_f^{\prime \prime }}
\exp \left[\i (\delta _{l_f^\prime}
-\delta _{l_f^{\prime \prime }})\right]
\sqrt{[l_f^\prime ][L^\prime ][l_f^{\prime \prime}]
[L^{\prime \prime }]} \nonumber \\
\times 
\left(
\begin{array}{ccc}
l_f^\prime & 2j & l_f^{\prime \prime} \\
0 & 0 & 0
\end{array}
\right)
\left( 
\begin{array}{ccc}
L^\prime  & 2j & L^{\prime \prime } \\
0 & 0 & 0
\end{array}
\right)
\left\{ 
\begin{array}{ccc}
L^\prime & L^{\prime \prime } & 2j \\
l_f^{\prime \prime} & l_f^\prime & l_0
\end{array}
\right\}
A_{l_f^\prime L^\prime }A^*_{l_f^{\prime \prime} L^{\prime \prime }}
\Biggr]
\end{eqnarray}
where $[l]\equiv 2l+1$ and $\beta _0=1$, so that the photoelectron angular
distribution after a two-photon detachment is characterized by
$\beta _2$ and $\beta _4$.

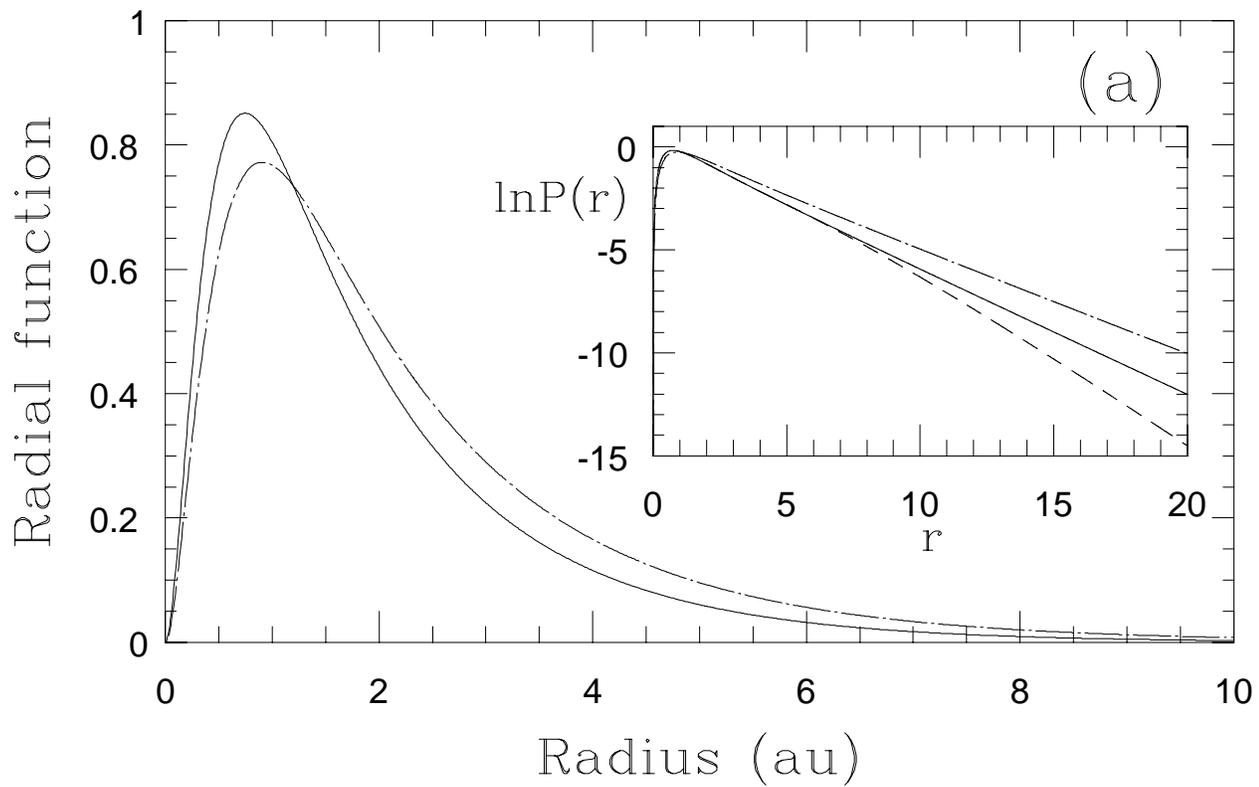
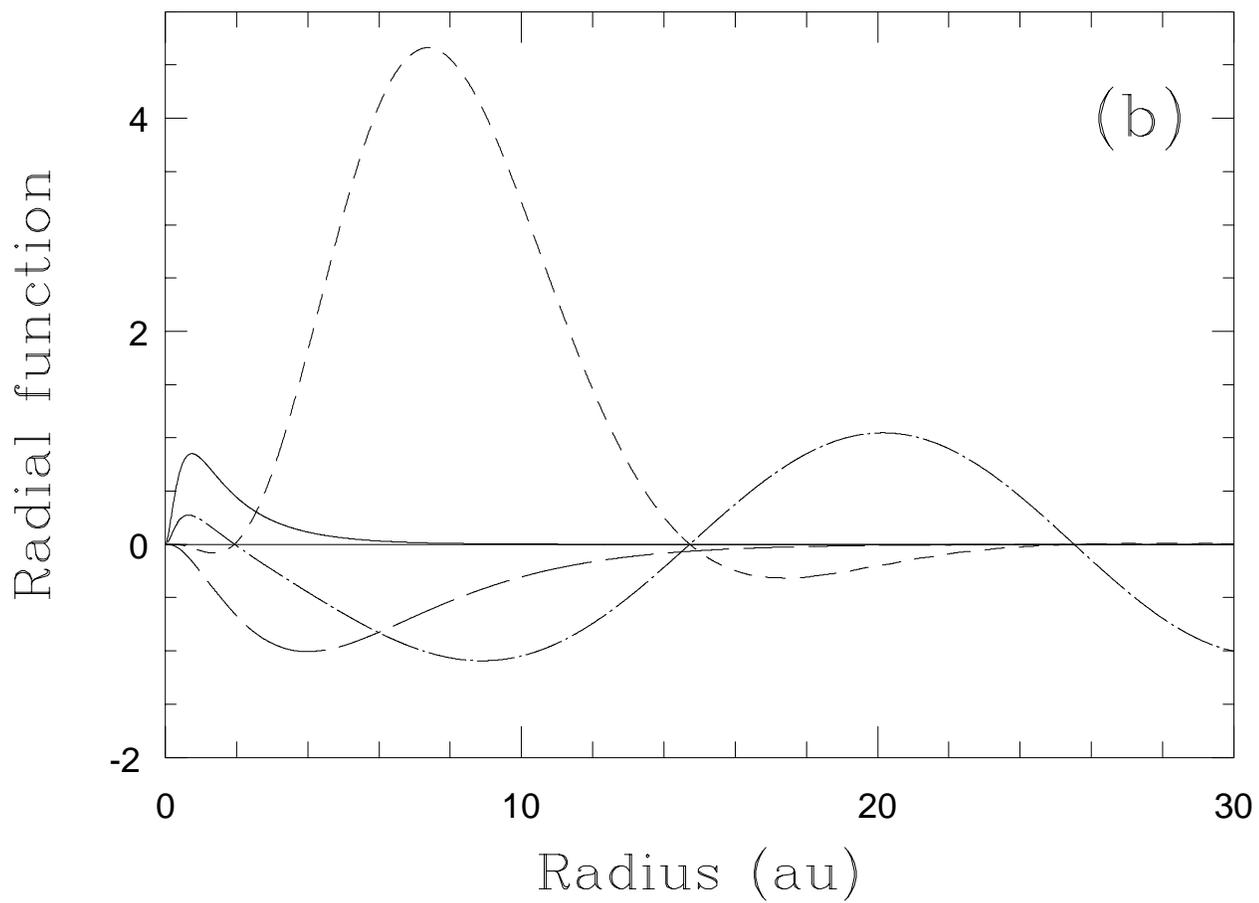
\begin{figure}
\caption{Wavefunctions of the F$^-$ ground state, effective intermediate
state and final state of the photoelectron. (a) Radial wavefunction of
the $2p$ subshell of F$^-$ in the HF approximation (\full ~,
$E_{2p}^{\rm HF}=-0.362$ Ryd), and that with a model potential added to
reproduce the experimental energy (\chain ~, $E_{2p}=-0.250$ Ryd).
The inset shows the same wavefunctions on the logarithmic scale, together
with the radial Roothaan-HF $2p$ radial wavefunction of F$^-$ from
Clementi and Roetti (1974), \dashed .
(b) \full ~, HF $2p$ wavefunction; \longbroken ~, effective wavefunction
$P_\omega (r)$, equation (\protect\ref{sh}), of the intermediate $d$
state at $\omega =0.226$ Ryd; \chain ~, final state $p$ ($^1D$)
wavefunction, $\varepsilon =0.09$ Ryd; \dashed ~, integrand of equation
(\protect\ref{dme1}) for the 2-photon amplitude $M_{pd}$.
\label{functions}}
\end{figure}

The self-consistent HF calculation of the F$^{-}$ ground state yields the
$2p$-electron energy $E_{2p}^{\rm HF}=-0.362$ Ryd, which is much lower than
its true value equal to the negative of the experimental electron affinity
of F: $E_{2p}^{\rm exp }=-0.250$ Ryd (Radtzig and Smirnov 1986). It is
often assumed that the HF radial wavefunction is still a good starting
point for calculations of multiphoton detachment, if the experimental
binding
energy is used in lieu of the HF value (Crance 1987a,b, Pan \etal 1990).
Pan \etal (1990) showed that the two-photon detachment cross sections
obtained with the dipole operator in the velocity form are very sensitive
to the $2p$-electron energy, while the length form results change little
when the HF energy is replaced by the experimental one. However, one should
use not only use the correct energy but, which is much more important,
the bound state wavefunction with the correct asymptotic
behaviour\footnote{The need for an asymptotically correct wavefunction is
clearly
illustrated by the adiabatic hyperspherical calculation of multiphoton
detachment from H$^-$ by Liu \etal (1992), where a 3.4\% change of
$\kappa $ results in a 25\% change of the 2-photon cross section.}. The
importance of large distances, where one can use the correct asymptotic
form of the bound-state wave function, speaks strongly in favour of using
the length form of the dipole operator (Gribakin and Kuchiev 1997a). To
correct the $2p$ wavefunction we solved the HF equations for the
F$^{-}$ ground state with an additional small
repulsive potential $V(r)=\alpha /[2(r^2+a^2)^2]$. We chose $\alpha =1$ and
$a=0.61$~au to ensure that the $2p$ energy was equal to the experimental
value. The HF and corrected $2p$ radial wavefunctions $P(r)=R(r)/r$ are
presented in figure \ref{functions}(a). The difference between them
appears to be small -- it does not exceed 10\% near the maximum. Their
asymptotic behaviour $P(r)\simeq A\exp (-\kappa r)$ corresponds to $A=0.94$
and 0.86, and  $\kappa =0.6$ and 0.5, respectively. The difference in
$\kappa $ means that the two wavefunctions are in fact quite different at
large distances.

The wavefunctions of the intermediate ($\nu l$) and final
($\varepsilon _fl_f$) states of the photoelectron are calculated in the HF
field of the frozen neutral $2p^5$ core. The photoelectron is coupled
to the core to form the total spin $S=0$ and angular momentum $L$: $L=1$
for the intermediate $l=0,\,2$ states, $L=0,\,2$ for $l_f=1$, and $L=2$
for $l_f=3$ final states. The intermediate states continua are discretized
and represented by a 70-state momentum grid with constant spacing
$\Delta p$.

There are two ways of calculating the two-photon amplitudes
$M_{{l_f}l}$ of equation (\ref{me2}). The first is by direct summation over
the intermediate states. It involves a non-trivial evaluation of the
free-free dipole matrix elements together with the accurate treatment of
pole- and $\delta $-type singularities (Korol 1994, 1997). Another way of
calculating such sums is by solving an inhomogeneous Schr\"odinger-type
equation for the effective radial function of the intermediate state
\begin{equation}\label{sh}
P_\omega (r) =\sum _\nu
\frac{P_{\nu l}(r)\langle \nu l\| \hat d\| n_0l_0\rangle }
{E_0 +\omega -E_\nu +\i 0}
\end{equation}
(Sternheimer 1951, Dalgarno and Lewis 1955). This wavefunction describes
the amplitude of finding the electron at different distances from the atom
after absorption of the first quantum. After calculation of
$P_\omega (r)$ the amplitude is obtained from the radial integral as
\begin{equation}\label{dme1}
M_{l_fl}( \omega ) =(-1)^{l_>}\sqrt{l_>}\int P_{\varepsilon _fl_f}( r)
P_\omega (r) r\d r ~,\qquad \left( l_>=\mbox{max}\{ l_f,l\}\right) ,
\end{equation}
In the present work we calculate the two-photon amplitudes using both
techniques. The second one is especially simple below the single-photon
detachment threshold, $\omega \lesssim |E_0|$, where $P_\omega (r)$ drops
exponentially at large distances:
$P_\omega (r)\propto \exp ( -K_\omega r ) $, where
$K_\omega =[2(|E_0| -\omega)]^{1/2}$. At finite distances $r<1/ \Delta p$
it can be computed easily by direct numerical summation in equation
(\ref{sh}).

It is instructive to look at the shape of the effective wavefunction
$P_\omega (r)$. As an example, figure \ref{functions}(b) shows this
function calculated
for the intermediate $d$ electron at $\omega =0.226$ Ryd for the HF ground
state. The maximum of $P_\omega (r)$ is shifted towards large radii,
compared to the maximum of the ground state wavefunction. Also shown
in figure \ref{functions}(b) are the radial wavefunction of the final
$p$-wave electron
($\varepsilon _f=2\omega + E_{2p}^{\rm HF}=0.09$ Ryd) and the integrand
$P_{\varepsilon _fl_f}(r)P_\omega (r) r$ of equation (\ref{dme1}). These
plots illustrate the point that the two-photon amplitude
$M_{l_fl_n}\left( \omega \right) $ is indeed determined by large
electron-atom separations (\ref{large}). Accordingly, the correct
asymptotic behaviour of the ground state wavefunction is crucial.

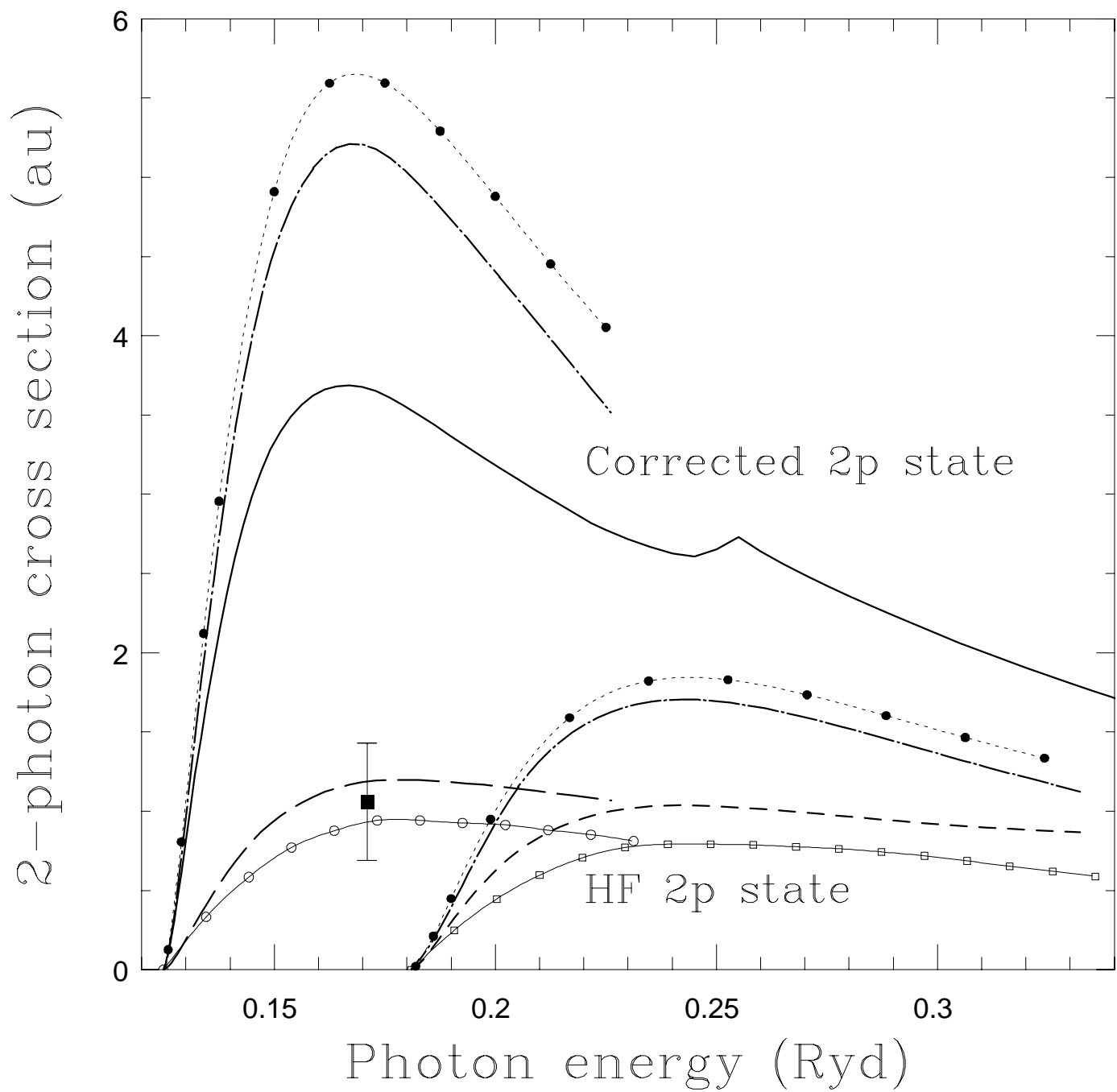
\begin{figure}
\caption{2-photon detachment cross sections. Present calculations:
\dashed , HF wavefunctions of the $2p$, intermediate and final states;
\full , same with the corrected $2p$ wavefunction; \chain , using plane
waves in the intermediate and final state; \longbroken , HF wave functions
combined with the experimental $2p$ energy; \protect\fullcirc , adiabatic
theory (equation (5) of Gribakin and Kuchiev 1997b). Other results:
$\protect\opensqr $ and $\protect\opencirc $, calculations by Pan \etal
(1990) with the HF and experimental binding energies, respectively;
$\protect\fullsqr $, experiment (Kwon \etal 1989). \label{crosssec}}
\end{figure}

In this work we compare the cross sections and angular asymmetry parameters
calculated in different approximations with both the HF and corrected $2p$ 
wavefunctions (figure \ref{functions}(a)). Let us first discuss the results
obtained with the HF energy of the $2p$ electron. It corresponds to the
two-photon threshold $\omega =0.181$ Ryd. The cross section calculated
from equations (\ref{cr})--(\ref{me2}) using the HF functions of the
initial, intermediate and final states are shown in figure \ref{crosssec}
by a short-dashed line. It is very similar to the dipole length
lowest-order HF results of Pan \etal (1990), although the latter
are about 10\% lower than ours. What is the source of this discrepancy?
Pan \etal  used the Roothaan-HF expansion of the bound state. This form of
the bound-state wavefunction has an incorrect asymptotic behaviour at
$r>7$, see figure \ref{functions}(a), inset. Because of the importance of
large electron-atom separations in the multiphoton processes even a small
error in the wave function could lead to some inaccuracies in the 2-photon
detachment amplitudes. In the work of Pan \etal electron correlation
effects were calculated. It was shown that they suppress the cross section
in F$^-$ by about 20\% at the maximum. These results shown in figure
\ref{crosssec} by open squares are still close to the HF curve.

The asymptotic parameters of the HF $2p$ wave function are $\kappa =0.6016$
and $A=0.94$. We use them in the adiabatic theory formula (equation (5)
of Gribakin and Kuchiev 1997b) and obtain the cross section shown in
figure \ref{crosssec} by solid dots. It reproduces the energy dependence
of the HF cross section well, though overestimates its magnitude by a
factor of two. This is a reasonable result, since the adiabatic
theory should only be valid for $n\gg 1$.
The calculations of Gribakin and Kuchiev (1997a) showed that for H$^-$
and $n=3$ the analytical adiabatic results are already 20\% accurate.
There are two approximations made in the adiabatic theory: (i) the use of
the Volkov wavefunction to describe the photoelectron, and (ii) the
saddle-point
calculation of the integral over time, which enables one to express the
amplitude in terms of the asymptotic parameters of the bound state.
In the weak-field regime the use of the Volkov function is equivalent
to the the so-called `free-electron' approximation (examined earlier by
Crance). In this approximation the wavefunctions of the photoelectron in
the intermediate and final states are described by plane waves. When we
do such calculation for F$^-$ (chain curve in figure \ref{crosssec}) the
results turn out to be very close to those of the adiabatic theory. This
means that the approximation (ii) of the adiabatic theory is in fact quite
good even at $n=2$.

When we use the experimental energy of the $2p$ electron together with
the HF wavefunctions the magnitude of the 2-photon cross section changes
very little (dashed line in figure \ref{crosssec}), as seen earlier by
Pan \etal (1990) for both HF and correlated results (open circles).
The HF results of Crance (1987a) are close to the above, and the cross
section of van der Hart (1996) is also similar, with a maximum of
1.27 au at $\omega =0.166$ Ryd.

However, when we use the corrected $2p$ wavefunction,
the photodetachment cross section increases more than three times. It is
shown by solid line in figure \ref{crosssec}, and we consider this to be
the best evaluation of the cross section for F$^-$. The cusp on the curve
corresponds to the single-photon threshold\footnote{This feature is a
consequence of the
Wigner threshold dependence $\sigma \propto \sqrt{\omega -E_0}$ of the
$s$-wave single-photon detachment from F$^-$.}. The same increase is also
demonstrated by the adiabatic theory (with modified asymptotic parameters
$\kappa =0.4998$ and $A=0.86$) and the plane-wave results. As we explained
earlier, this ``surprising'' sensitivity of the multiphoton detachment
probabilities to the asymptotic form of the bound-state wavefunction is a
direct consequence of the dominant role of large electron-atom separations
in this problem (Gribakin and Kuchiev 1997a).

The error induced by the use of an asymptotically incorrect wave
function can be estimated within the adiabatic approach. It turns out that
if one uses the experimental binding energy $E_0=-\kappa ^2/2$ together
with an incorrect bound state $P(r)\propto \exp (-\kappa 'r)$, the
$n$-photon cross section acquires an error factor
\begin{equation}\label{err}
\left( 1-\sqrt{\frac{\pi}{2}}\frac{\Delta \kappa }
{\sqrt{\omega }}\right) ^2~,
\end{equation}
where $\Delta \kappa =\kappa '-\kappa $. This equation implies
that the relative error in the amplitude is $\sim \Delta \kappa R$, where
$R=1/\sqrt{\omega }$ is the large radius from equation (\ref{large}).
For $\kappa '>\kappa $ the error factor is smaller than unity. Thus,
using a stronger bound wave function leads to an underestimation of the
cross section. For F$^-$ the factor (\ref{err}) calculated for
$\Delta \kappa =0.1$ and $\omega =0.085$ au near the cross section maximum,
yields 0.33. This value agrees with the difference between the cross
sections observed in figure \ref{crosssec}. The only other work
that used an asymptotically correct $2p$ wave function was the model
potential calculation of Robinson and Geltman (1967), which produced
a cross section two times greater than those of Crance, Pan \etal and
van der Hart.

To estimate the size of possible errors introduced by our way of correcting
the $2p$ wavefunction we have examined the dependence of our cross section
on the choice of $\alpha $ and $a$ in the repulsive
potential. We find that as long as the asymptotic behaviour of the $2p$
state remains correct, the two-photon cross sections are always
proportional to that shown by the solid line in figure \ref{crosssec}.
Different pairs of $\alpha $ and $a$ result in the variation of $A$,
and the magnitude of the cross section is simply proportional to $A^2$. Our
value of $A=0.86$ is close to $A=0.84$ from Radtsig and Smirnov (1986) and
we are sure that our results are basically correct. Even a large 10\%
uncertainty in the value of $A$ would mean a maximal 20\% error in the cross
section. In any case the cross section will be much larger than those
obtained with the HF $2p$ ground state.


The difference between experimental and HF values of the $2p$ energy is
a manifestation of electron correlations. It influences the result via
the asymptotic behaviour of the ground-state wavefunction. This is by far
the most important correlation effect in multiphoton detachment.
The use of the asymptotically correct $2p$ wavefunction changes the
cross section by a factor of three, which is much greater than other
correlation effects (Pan \etal 1990). This fact distinguishes this problem
from the single-photon processes, where other correlation effects are
essential.

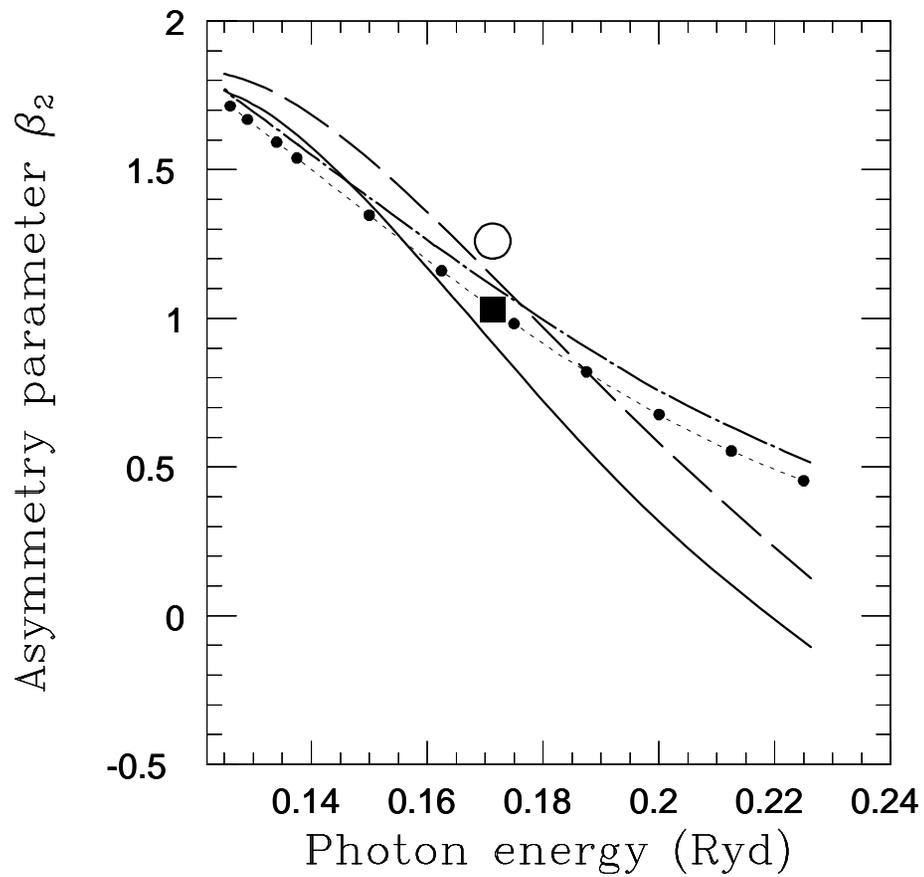
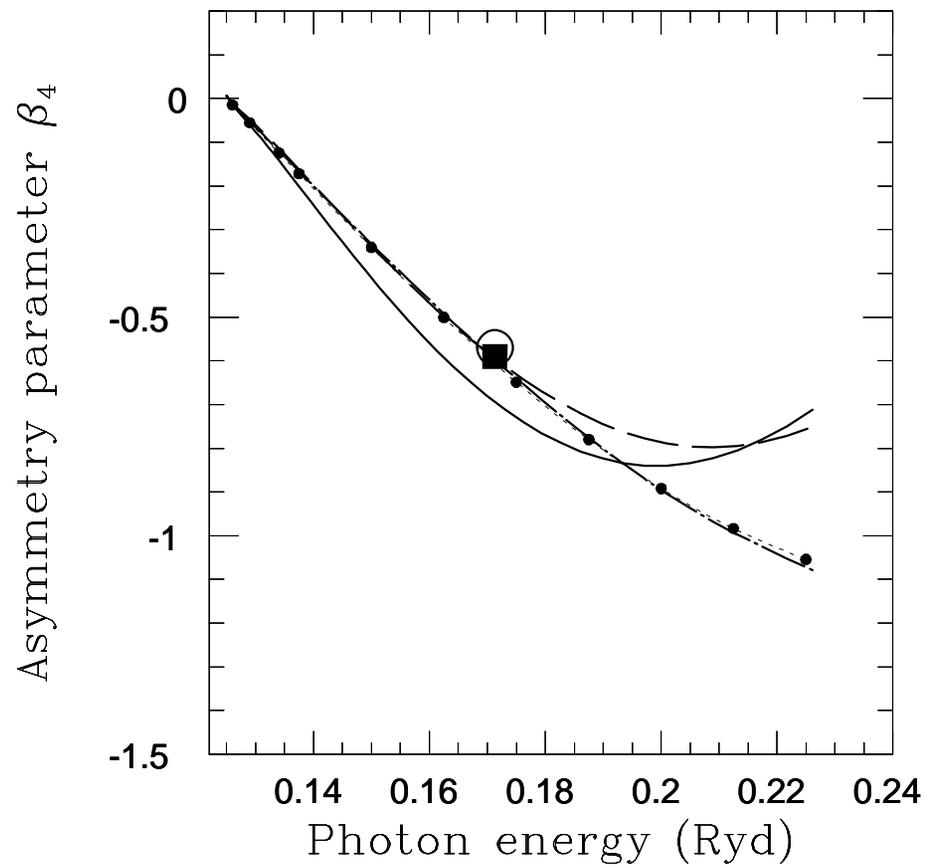
\begin{figure}
\caption{Photoelectron angular distribution parameters.
Present calculation: \longbroken , HF wavefunctions of the initial,
intermediate and final states, experimental $2p$ energy; \full ,
corrected $2p$ wavefunction, HF intermediate and final states; \chain ,
same with the plane wave in the intermediate and final states;
\protect\fullcirc , $\beta $ parameters obtained from the adiabatic theory
(equations (3), (4) of Gribakin and Kuchiev 1997b). Other results:
$\protect\opencirc $, correlated length results by Pan and Starace
(1991); $\protect\fullsqr $, experiment (Blondel and Delsart 1993).
\label{angasym}}
\end{figure}

The angular asymmetry parameters $\beta _2$ and $\beta _4$ calculated
using the experimental $2p$ energy are shown in figure \ref{angasym},
together with the correlated length results of Pan and Starace (1991)
and experimental points of Blondel and Delsart (1993) at $\omega =0.171$
Ryd. The asymmetry parameters (\ref{beta}) are relative quantities, and
the results of different calculations are much closer for them than 
for the absolute values of the photodetachment cross sections.
The adiabatic theory is again in good agreement with the plane-wave
approximation, especially in $\beta _4$. It appears that this parameter
is on the whole less sensitive to the details of the calculation,
because it is simply proportional to the amplitude of $f$ wave emission,
and there is no interference in the sum in equation (\ref{beta}) for 
$\beta _4$. The
experimental values of $\beta $ for F$^-$ obtained in the earlier work of
Blondel \etal (1992) are close to those of Blondel and Delsart (1993).
This is why F$^-$ serves as a good benchmark for angular asymmetry
calculations. The perfect agreement between adiabatic theory and the
experiment is probably fortuitous. Figure \ref{angasym} indicates that for
the experiment to be able to distinguish between various theoretical data
one would wish to make measurements at higher photon energies, where
the results of different approximations diverge.

From the theoretical point of view it seems that the total cross sections
and the angular asymmetry parameters are determined by different
physical features of the problem. The absolute size of the cross sections
is very sensitive to the asymptotic behaviour of the bound state wave
function. This sensitivity {\it increases} for large-$n$ processes, when
$\omega $ become smaller, as suggested by estimate (\ref{err}). The cross
sections also depend on the atomic potential which acts on the
photoelectron, hence the difference between the results obtained with
the HF and plane waves. This latter effect is {\it suppressed} for larger
$n$, since this re-scattering of the photoelectron is inversely
proportional to some power of large $R$. The angular asymmetry parameters
are affected by the electron-atom potential via the scattering phaseshifts
$\delta _{l_f}$. However, for large $n$ and small photoelectron  energy
$E\sim \omega $ the phaseshifts should be close to integer multiples of
$\pi $. Besides that, contributions of higher partial waves become
dominant. They are almost unaffected by the atomic potential and the
adiabatic theory should become very accurate.

In summary, we have shown by direct numerical calculations that in
agreement with the adiabatic theory, the multiphoton detachment rates are
very sensitive to the asymptotic behaviour of the bound state wavefunction.
For fluorine this means that the true 2-photon detachment cross sections
are substantially higher than it was believed earlier. The discrepancy
revealed is much greater than other electron correlation effects.

\ack
This work was supported by the Australian Research Council.
One of us (VKI) would like to acknowledge the hospitality extended to him
at the School of Physics at the University of New South Wales, and the
support of his visit by the Gordon Godfrey fund.

\section*{References}

\begin{harvard}

\item[] Blondel C, Crance M, Delsart C and Giraud A 1992
{\it J. Physique II} {\bf 2} 839

\item[] Blondel C and Delsart C 1993 {\it Nucl. Instrum. Methods B}
{\bf 79} 156

\item[] Clementi E and Roetti C 1974 {\it At. Data Nucl. Data Tables}
{\bf 14} 177

\item[]Crance M 1987a {\it J. Phys. B: At. Mol. Phys.} {\bf 20} L411

\item[]\dash 1987b {\it J. Phys. B: At. Mol. Phys.} {\bf 20} 6553

\item[] Dalgarno A and Lewis J T 1955 {\it Proc. R. Soc. (London)}
{\bf A233} 70

\item[] Gribakin G F and Kuchiev M Yu 1997a {\it Phys. Rev. A} {\bf 55}
3760

\item[] Gribakin G F and Kuchiev M Yu 1997b {\it J. Phys. B:
At. Mol. Opt. Phys.} {\bf 30} L657

\item[] Keldysh L V 1964 {\em Zh. Eksp. Teor. Fiz.} {\bf 47} 1945
[1965 {\em Sov. Phys. JETP} {\bf 20} 1307].

\item[] Korol A V 1994 {\it J. Phys. B: At. Mol. Phys.} {\bf 27} 155

\item[] Korol A V 1997 unpublished

\item[] Kuchiev M Yu and Ostrovsky V N 1998 {\it J. Phys. B:
At. Mol. Opt. Phys.} {\bf 31} to appear

\item[] Kwon N, Armstrong P S, Olsson T, Trainham R and Larson D J 1989
{\it Phys. Rev. A} {\bf 40} 676

\item[] Liu C-R, Gao B and Starace A F 1992 {\it Phys. Rev. A} {\bf 46}
5985

\item[] Pan C, Gao B and Starace A F 1990 {\it Phys. Rev. A} {\bf 41} 6271

\item[] Pan C and Starace A F 1991 {\it Phys. Rev. A} {\bf 44} 324

\item[] Radtsig A A and Smirnov B M 1986 {\it Parameters of Atoms and
Atomic Ions} (Moscow: Energoatomizdat)

\item[] Robinson E J and Geltman S 1967 {\it Phys. Rev.} {\bf 153} 4

\item[] Sternheimer R M 1951 {\it Phys. Rev.} {\bf 84} 244

\item[] van der Hart H W 1996 {\it J. Phys. B: At. Mol. Phys.} {\bf 29}
3059

\end{harvard}

\Figures

\end{document}